\documentclass[oldversion]{aa}
\usepackage{txfonts}
\usepackage{epsfig}
\usepackage{graphicx}
\author{Zapata et al.}
\begin{document}

\titlerunning{Circumbinary Molecular Rings Around Young Stars}
\authorrunning{Zapata et al.}

\title{Circumbinary Molecular Rings Around Young Stars in Orion}

\author{Luis A. Zapata\inst{1,2,3}, Paul T. P. Ho\inst{3,4},
Luis F. Rodr\'\i guez\inst{2}, Peter Schilke\inst{1}, and
Stan Kurtz\inst{2}}

\institute{Max-Planck-Institut f\"{u}r Radioastronomie, Auf dem H\"ugel 
69,
53121, Bonn, Germany
\and CRyA, Universidad Nacional Aut\'onoma de M\'exico,
Apdo. Postal 3-72 (Xangari), 58089 Morelia, Michoac\'an, M\'exico
\and Harvard-Smithsonian Center for Astrophysics, 60 Garden Street,
Cambridge, MA 02138, USA
\and Academia Sinica Institute of Astronomy and Astrophysics,
Taipei, Taiwan}

\date{Received -- / Accepted --}

\offprints{Luis Zapata, \email{lzapata@mpifr-bonn.mpg.de}}

\abstract{We present high angular resolution 1.3~mm continuum, methyl 
cyanide
molecular line, and 7~mm continuum observations made with the
Submillimeter Array and the Very Large Array, toward the most highly 
obscured
and southern part of the massive star forming region OMC1S located behind 
the
Orion Nebula. We find two flattened and rotating molecular structures with 
sizes of a few hundred
astronomical units suggestive of circumbinary molecular rings produced by 
the presence
of two stars with very compact circumstellar disks with sizes and separations of
about 50 AU, associated with the young stellar objects 139-409 and 
134-411.
Furthermore, these two circumbinary rotating rings are related to two 
compact and bright
{\it hot molecular cores}. The dynamic mass of the binary systems obtained from 
our data
are $\geq$ 4 M$_\odot$ for 139-409 and $\geq$ 0.5 M$_\odot$ for 134-411.
This result supports the idea that intermediate-mass stars will form 
through
{\it circumstellar disks} and jets/outflows, as the low mass stars do.
Furthermore, when intermediate-mass stars are in multiple systems
they seem to form a circumbinary ring similar to those seen
in young, multiple low-mass systems (e.g., GG Tau and UY Aur).}

\keywords{
stars: pre-main sequence  --
ISM: jets and outflows --
ISM: individual: (Orion-S, OMC1-S, Orion South, M42) --
ISM: Molecules, Radio Lines --
ISM: Circumstellar Disks --
ISM: Binary stars --
ISM: Circumbinary Disks --}

\maketitle

\section{Introduction}
About thirty years ago a new major puzzle emerged in the
field of stellar astrophysics:
How do massive stars form?, where massive stars
are those with more than $10$ M$_{\odot}$
(Kahn 1974; Larson \& Starrfield 1971; Yorke \& Kruegel 1977).
It was believed that the powerful radiation fields and
stellar winds produced at the very beginning of
their lives will inhibit accretion of material
limiting their mass growth to about $10$ M$_{\odot}$.
In the two last decades, several alternatives
have been proposed to solve this puzzle.
Among the most important are the formation of massive stars through
dense disks and jets/outflows (Nakano 1989; Jijina \& Adams 1996),
in very dense and turbulent cores (McKee \& Tan 2002),
through the merging of smaller stars (Bonnell et al. 1998),
and through ionized accretion flows that forms an ionized disk or
torus around a group of stars (Sollins et al. 2005; Keto \& Wood 2006).
Discriminating between these alternatives
remains an observational challenge (Cesaroni et al. 2007).

OMC1S or Orion-S is the ``twin'' dusty massive molecular
core of the Orion BN-KL core. It is located almost at the same angular distance from
the  ``Trapezium'' as Orion BN-KL ($\sim$ 1$'$), but to the 
southwest of the former.
We adopt a distance of 460 pc to the Orion Nebula (Bally et al. 2000).
The OMC1S region has a mass of about 100 M$_{\odot}$, similar to that 
reported for BN-
KL (Mezger at al. 1990), but with a bolometric luminosity of $\sim$ 10$^4$ 
L$_{\odot}$,
which is a factor of 10 less (Mezger at al. 1990; Drapatz et al. 1983).
This difference in luminosity might be attributed to OMC1S
being less evolved
than Orion BN-KL, as inferred if one compares the molecular
line emission from both regions (McMullin et al. 1993).
With time, the massive stars forming in OMC1S might reach their final
masses and shine with much larger luminosity than now.
This possible evolutionary scheme has been also suggested to be taking 
place
in the NGC6334I region (Rodr\'\i guez, Zapata \& Ho 2007).

In this letter we report for the first time the possible presence of
two circumbinary molecular rotating rings located in the OMC1S region
with sizes of a few hundred Astronomical Units (AU) around two very 
compact
circumstellar disks and that are associated with intermediate-mass 
(proto)stars.

\begin{figure*}[ht]
\begin{center}\hspace{-0.12cm}
\includegraphics[width=12.6 cm]{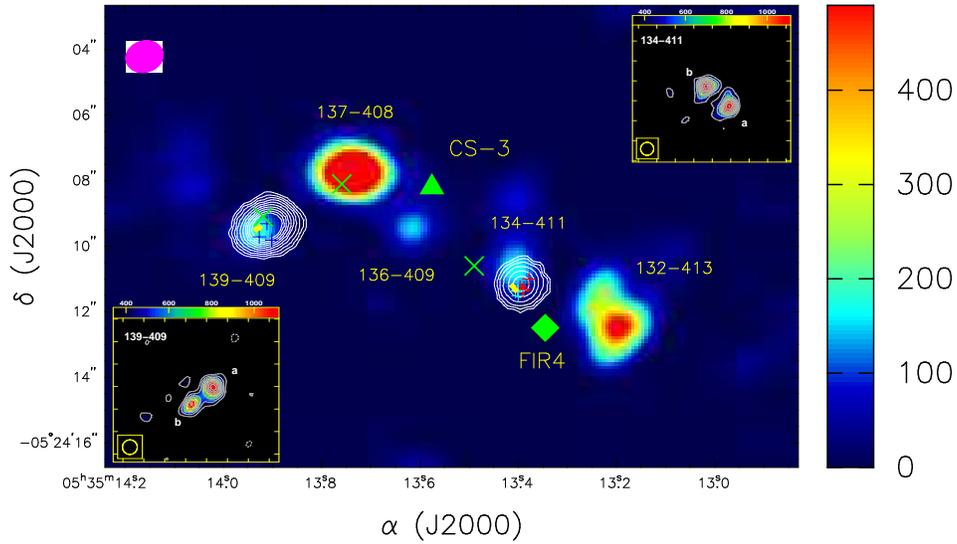}
\caption{\scriptsize SMA 1.3 mm continuum color image of the southern most
region of OMC1S (from Zapata et al. 2005),
overlaying on the CH$_3$CN[12$_4$-11$_4$] integrated molecular emission of 
the {\it hot molecular cores} 139-409
and 134-411 (white contours). The contours of the CH$_3$CN[12$_4$-11$_4$] 
integrated molecular emission
are -5, 5, 10, 15, 20, 25, 30, 35, 40, 45 and 50 times 120 mJy beam$^{-1}$
km s$^{-1}$, the rms noise of the image.
The integration is over a velocity range of -5 to 17 km s$^{-1}$.
The millimeter sources 139-409, 134-411, 132-413, 135-409 and 137-408
were reported for the first time by Zapata et al. (2005).
The synthesized beam of the CH$_3$CN image is 1.13$''$ $\times$ 0.93$''$ 
with a P.A. = -73$^\circ$ and
it is shown in the upper left corner.
The scale bar indicates the 1.3 mm continuum emission in mJy beam$^{-1}$.
The yellow rhombi indicate the positions of the 7 mm continuum compact 
radio binaries.
The green rhombus and triangle denote the position of the source FIR 4 
(Mezger et al. 1990)
and the millimeter source CS 3 (Mundy et al. 1986), respectively.
The blue and red crosses indicate the position of the blue- and 
red-shifted
H$_2$O maser spots, respectively, reported by Gaume et al. (1998).
Note that the masers associated with the hot molecular core (134-411) show a large
velocity gradient, going from -20 to +45 km s$^{-1}$.
The green 
`X' symbols indicate
the position of the 3 mm BIMA continuum sources reported  by Eisner \& 
Carpenter (2006).
In the left bottom and right upper corners we show the 7 mm continuum 
compact radio binaries located
on the centers of the hot cores 139-409 and 134-411.
The scale bar indicates the 7 mm continuum emission in 10$^{-3}$ mJy 
beam$^{-1}$
on both images. In the left bottom corner box the
contours are -2, 2, 3, 4, 5, 6, 7, 8, 9 times 0.15 mJy beam$^{-1}$, the 
rms
noise of the image. In the right upper corner box the contours are -2, 2, 
3, 4, 5, 6, 7, 8, 9
times 0.15 mJy beam$^{-1}$ the rms noise of the image. The synthesized 
beam is shown
in the bottom left corner of each box.  The sizes of these boxes are 
0.6$''$ $\times$ 0.6$''$. }
\label{fig1}
\end{center}
\end{figure*}

\section{Observations}

The observations were made with the Submillimeter Array (SMA)\footnote{The 
Submillimeter Array
is a joint project between the Smithsonian Astrophysical Observatory and
the Academia Sinica Institute of Astronomy and Astrophysics, and is
funded by the Smithsonian Institution and the Academia Sinica.} and the 
Very Large
Array (VLA)\footnote{The National Radio
Astronomy Observatory is a facility of the National Science Foundation
operated under cooperative agreement by Associated Universities, Inc.}
during 2004 September 2 and November 10, respectively.

The SMA was in its extended configuration,
which includes 21 independent baselines ranging in projected length from 
16 to 180 m.
The phase reference center of the observations was
R.A. = 05$^h$35$^m$14$^s$, decl.= -05$^\circ$24$'$00$''$ (J2000.0).
The receivers were tuned at a frequency of 230.534 GHz in the upper 
sideband (USB),
while the lower sideband (LSB) was centered at 220.53 GHz.
The CH$_3$CN[12$_4$-11$_4$]  transition was detected in the LSB at a frequency of 
220.679 GHz\footnote{ Several hyperfine transitions of the 
CH$_3$CN[12-11] molecule were detected toward these sources, however, the
images were made from this transition.}.
The full bandwidth of the SMA correlator is 4 GHz (2 GHz in each band).
The SMA digital correlator was configured in spectral windows
(``chunks'') of 104 MHz each, with 32 channels distributed
over each spectral window, providing a resolution of
3.25 MHz (4.29 km s$^{-1}$) per channel.

The zenith opacity measured ($\tau_{230 GHz}$)
with the NRAO tipping radiometer located at the Caltech Submillimeter 
Observatory was $\sim$ 0.04,
indicating very good weather conditions during the experiment. 
Observations of Callisto
provided the absolute scale for the flux density calibration.
The phase and amplitude calibrators were the quasars 0423-013 and 3C 120, 
with measured flux densities
of 2.343 $\pm$ 0.006 and 0.563 $\pm$ 0.003 Jy, respectively.
The absolute flux density calibration uncertainty is estimated
to be 20\%, based on SMA monitoring of quasars.
Further technical descriptions of the SMA and its calibration
schemes are found in Ho et al. (2004).

The VLA was in its most extended A-configuration,
providing baselines with a maximum projected length of 36 km.
The central frequency
observed was 43.34 GHz. The absolute amplitude calibrator was 1331+305
(with an adopted flux density of 1.45 Jy)
and the phase calibrator was 0541-056 (with a bootstrapped flux density of 
1.10 $\pm$ 0.09).
The phase center of these observations was the same as for the SMA 
observations.

The SMA data were calibrated using the MIR\footnote{The MIR cookbook by C. 
Qi can be found
at http://cfa-www.harvard.edu/$\sim$cqi/mircook.html} software package, 
while the VLA data
were calibrated using the AIPS software. The calibrated data were imaged 
and analyzed
in the standard manner using the MIRIAD and AIPS packages.
We used the ROBUST parameter set to 0, for an optimal compromise between 
sensitivity and angular resolution.
The resulting continuum image rms at 7 mm was 0.15 mJy beam$^{-1}$, at an 
angular resolution
of $0\rlap.{''}05$ $\times$ $0\rlap.{''}04$ with a P.A. = -6$^\circ$. For the SMA 
observations
the line image rms noise was 120 mJy beam$^{-1}$ for each channel at an 
angular resolution of
$1\rlap.{''}13$ $\times$ $0\rlap.{''}93$ with a P.A. = -73$^\circ$.

\begin{figure*}[ht]
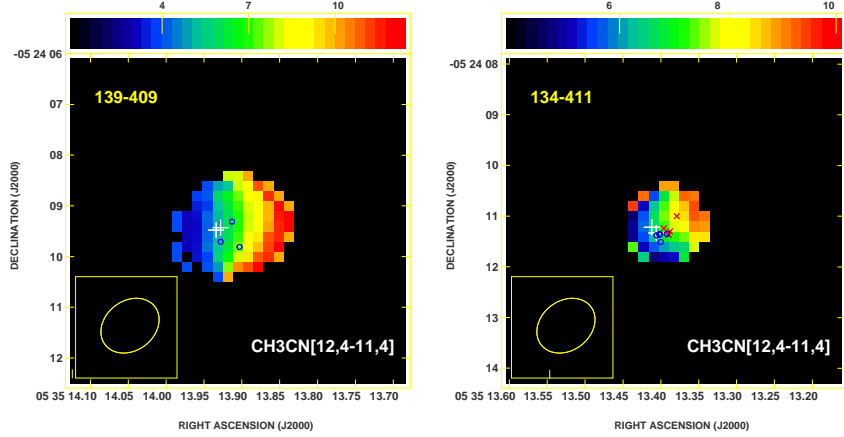

\begin{center}
\includegraphics[scale=.3]{7782f2.eps}
\includegraphics[scale=.3]{7782f3.eps}
\caption{\scriptsize CH$_3$CN[12$_4$-11$_4$] first moment
  maps of the molecular hot cores 139-409 and 134-411.
  The white crosses indicate the positions of the 7 mm compact radio 
binaries shown in Figure 1.
  The systemic LSR radial velocity of the ambient molecular cloud is about 
7 km s$^{-1}$.
  The scale bar indicates the LSR radial velocities of the core in km 
s$^{-1}$.
  Note that both cores show rotation of a few kilometers per second,
  with different orientations. The synthesized beam is shown
  in the bottom left corner of each image. The spectral resolution is 4 km 
s$^{-1}$.
  Note that our spectral resolution is comparable with the linewidth
  of the CH3CN transition. The blue circles and red crosses indicate the 
  position of the blue- and red-shifted H$_2$O masers, as in Figure 1.}
\end{center}
\label{fig2}
\end{figure*}

\section{Results and Discussion}

\subsection{1 mm Continuum and Molecular Emission}

In Figure \ref{fig1}, we have overlayed the SMA 1.3 mm continuum image 
from Zapata et al. (2005),
with the integrated CH$_3$CN[12$_4$-11$_4$] emission in the
southern, most obscured part of OMC1S. Furthermore, we have also included
on this image the positions of the water maser spots from Gaume et al. 
(1998),
the positions of the strong 3 mm continuum and FIR sources located here
(Mezger et al. 1990, Mundy et al. 1986, Eisner and Carpenter 2006), and 
the positions of the
7 mm binary systems associated with the sources 139-409 and 134-411.

In this image, we can see five 1.3 mm continuum
sources (139-409, 137-408, 136-409, 134-411,
and 132-413), but with only two sources (139-409 and 134-411) showing very
compact CH$_3$CN[12$_4$-11$_4$] emission.
Some of these continuum sources (139-409, 137-408, 134-411 and 132-413) 
are
the 1.3 mm counterparts of the 3.6 cm, 1.3 cm and 3 mm
sources reported by Eisner \& Carpenter (2006) and Zapata et al. (2004a, 
2004b) and which are
interpreted as UC HII regions, ionized thermal jets and/or massive circumstellar 
disks.
The 139-409 and 134-411 sources show strong compact molecular emission 
also in different lines
(e.g. series of CH$_3$CN, CH$_3$OH, SO$_2$, $^{34}$SO, H$_2$CO, etc), and
are thus related with two bright, very compact {\it hot molecular cores}
associated with intermediate-mass stars (Zapata et al., in prep.).

The CH$_3$CN[12$_4$-11$_4$] line molecular emission associated with the
source 139-409 has a deconvolved size of $0\rlap.{''}64 \pm  0\rlap.{''}03$  $\times$
$0\rlap.{''}45 \pm 0\rlap.{''}04$ (or 294 AU $\pm$ $14$ AU  $\times$ 207 AU $\pm$ $18$ AU)
with a P.A.of 87$^\circ$ $\pm$ 7$^\circ$, while the molecular emission
associated with the source 134-411 has dimensions of $0\rlap.{''}36 \pm 0\rlap.{''}07$
$\times$ $\leq 0\rlap.{''}33$ (or 166 AU $\pm$ $32$ AU  $\times$ $\leq$ 151 AU)
with a P.A.of 116$^\circ$ $\pm$ 10$^\circ$.

The CH$_3$CN[12$_4$-11$_4$] molecular emission shows a total velocity 
shift of 5 km s$^{-1}$ for the object 139-409 and of 2.5 km s$^{-1}$ for the object
134-411 and if one assumes that this molecular
gas is rotating in a Keplerian way and uses the deconvolved sizes of the 
molecular
structures presented above, we calculated a lower limit for the mass
of the central objects shown in Table 1.
Since these are lower limits, we think that the binaries at the
center of these circumstellar disks are formed by intermediate-mass stars.
The velocity gradients in both objects were determined from a 
simultaneous
Gaussian fitting to the {\it K}= 0 to 6 components of the CH$_3$CN[12-11]
transition.
Finally, the hypothesis of rotation is suggested by the observation 
that the major axis
of the molecular structures seems to be approximately aligned with 
the direction
of the velocity gradients.

\begin{center}
\begin{table}[h!]
\scriptsize
\caption{Physical Parameters of the Circumbinary Rings}
\begin{tabular}{lcccccc}
\hline \hline
        &         &         &Deconv.&    Rot.  &  \multicolumn{2}{c}{Mass} 
\\
\cline{6-7}
Name   & R.A.    & Dec.    &  Radius & Vel.    & Dyn. & Gas$^a$       \\
        & [J2000] & [J2000] &  [AU]   & [km s$^{-1}$] & [M$_{\odot}]$ & 
[M$_{\odot}]$ \\
\hline
139-409 & 05 35 13.912 & -05 24 09.40 & 147$\pm$ $7$& 5   & $\geq$4   & 
0.08\\
134-411 & 05 35 13.394 & -05 24 11.15 &  83$\pm$ $16$&2.5 & $\geq$0.5 & 
0.12\\
\hline \hline
\end{tabular}\\

The Circumstellar Disks\\

\begin{tabular}{lccccc}
\hline \hline
        &         &         & Flux     & Undeconv. &  Mass of the  \\
Name   & R.A.    & Dec.    & Density  & Radius &  Gas$^a$  \\
        & [J2000] & [J2000] & [mJy]    & [AU]   & [M$_{\odot}]$ \\
\hline
139-409a & 05 35 13.928 & -05 24 09.41 & 2.5 $\pm$0.5 & 25 & 0.05 \\
139-409b & 05 35 13.933 & -05 24 09.47 & 3.2 $\pm$0.5 & 20 & 0.06 \\
134-411a & 05 35 13.406 & -05 24 11.33 & 3.3 $\pm$0.5 & 25 & 0.06 \\
134-411b & 05 35 13.412 & -05 24 11.22 & 3.2 $\pm$0.5 & 25 & 0.06 \\
\hline
\end{tabular} \\

{\scriptsize (a): The masses of the gas were obtained assuming
a dust temperature value of 100 K, an adopted value
of $\kappa_{1.3 mm}$ = 1.5 cm$^2$ g$^{-1}$ (the average of the values
of 1.0 cm$^2$ g$^{-1}$, valid for grains with thick dust mantles,
and 2.0  cm$^2$ g$^{-1}$, valid for grains without mantles) and
the flux densities at 1.3 mm of 180 mJy for 139-409 and of 270 mJy for 134-411. }
\label{submmcont}
\end{table}
\end{center}

\subsection{7 mm Continuum Emission}

The sources 139-409 and 134-411 in the 7 mm continuum observations were 
resolved into two
compact binary systems with separations and sizes of about 50 AU (see 
Figure 1)
that are interpreted as very compact circumstellar disks.
The small size of these disks might be explained as tidally truncated 
disks as
those observed in the L1551 IRS5 system (Rodr\'\i guez et al. 1998).
 In the case of 134-411 the orientation of the H$_2$O masers is very 
close to that
observed for the circumbinary molecular disk (see Figure 2), suggesting 
that maybe these
masers are either associated with it. However, the velocity gradient
observed in the masers is too large to be explained in terms of
Keplerian rotation.

From the millimeter observations of Zapata et al. (2005) and the 7 mm 
continuum
observations presented here, we calculate that the sources 139-409 and 
134-411 have
average spectral indices of 2.6 $\pm$ 0.3 and  2.7 $\pm$ 0.3, 
respectively.
We then suggest that at 7 and 1.3 mm we are observing optically thin dust 
emission with a dust mass
opacity coefficient that varies with frequency as $\kappa \propto 
\nu^{0.6}$ and $\kappa \propto  \nu^{0.7}$
(suggesting grain growth). With this information we can estimate the 
enclosed masses of the 7 mm sources.
Assuming optically thin, isothermal dust emission and a gas-to-dust ratio 
of 100 (Sodroski et al. 1997)
and an adopted value of $\kappa_{1.3 mm}$ = 1.5 cm$^2$ g$^{-1}$
(and then $\kappa_{7 mm}$ equal to 0.46 for 139-409
and 0.54 cm$^2$ g$^{-1}$ for 134-411), and a typical dust temperature 
value of 100 K,
we derive total enclosed masses for those putative circumstellar disks 
that are shown in Table 1. From the results of this Table we conclude that
the masses of these systems are dominated by the stellar components, with the disks
contributing with masses of order 0.1 $M_\odot$. We also note that the mass associated
with the circumbinary disks is comparable with that associated with the
circumstellar disks (see Table 1).

\begin{figure}[ht]
\begin{center}
\includegraphics[scale=0.28]{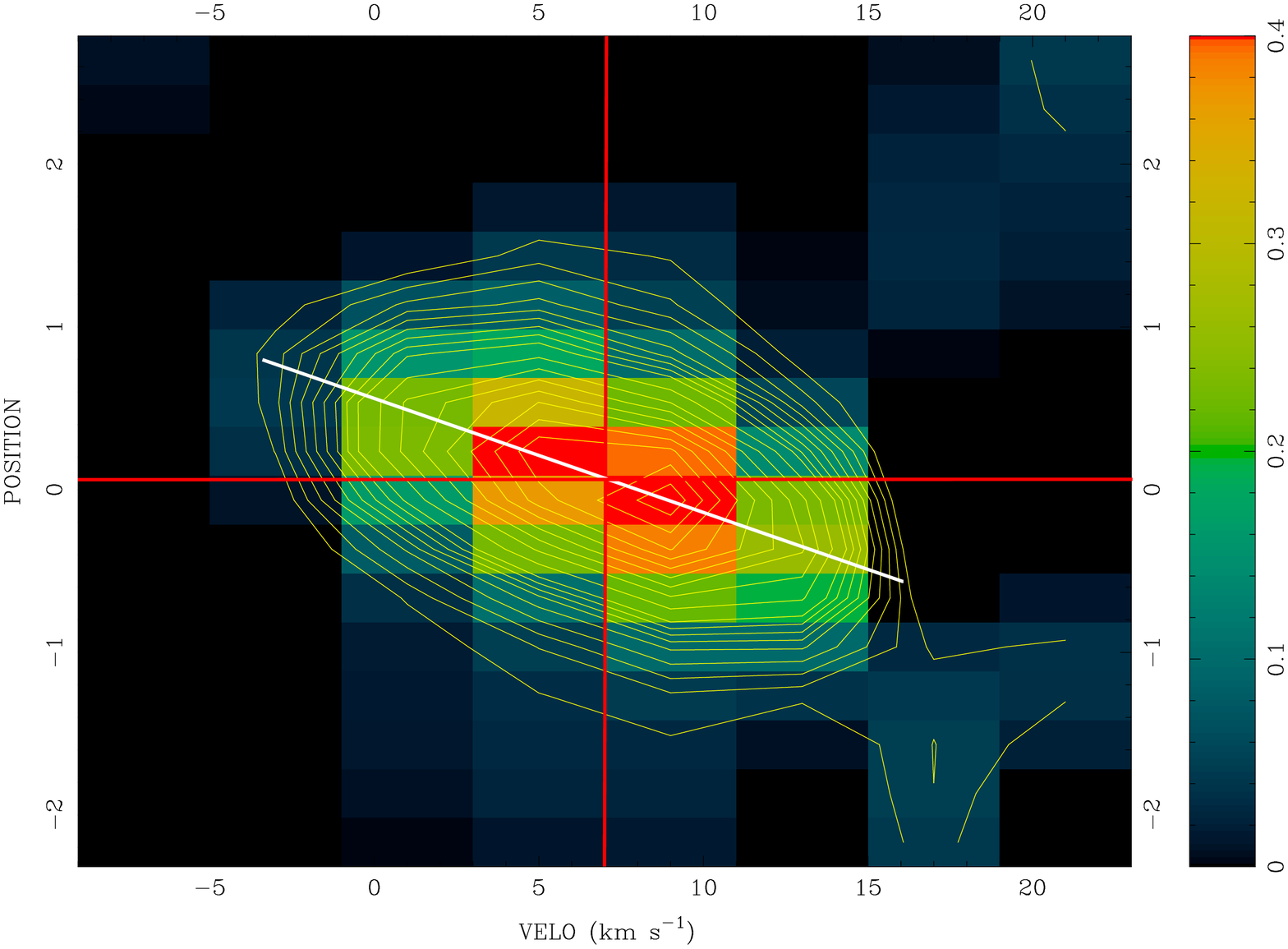}
\includegraphics[scale=0.28]{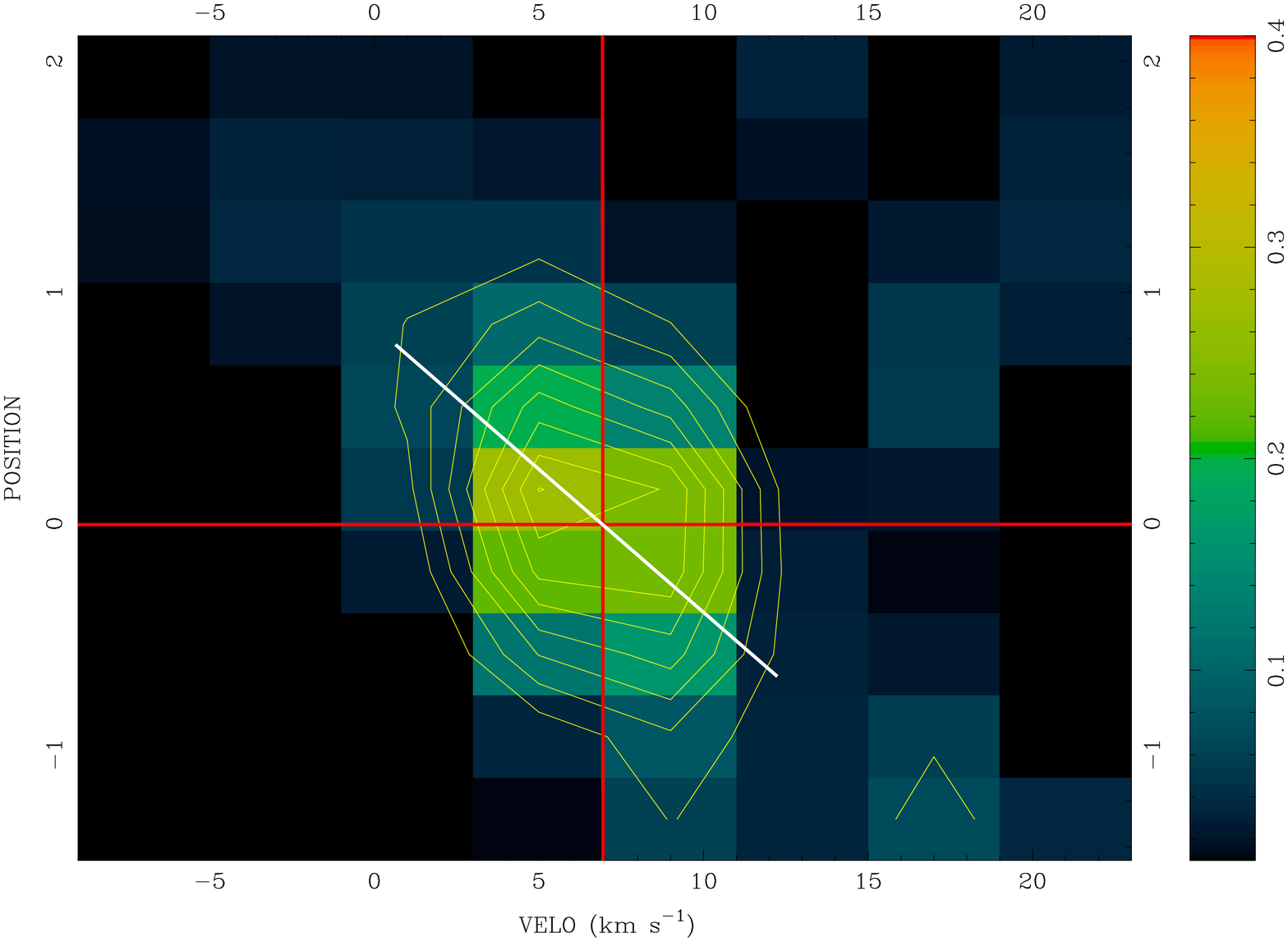}
\caption{\scriptsize Position-velocity diagrams of the 
CH$_3$CN[12$_4$-11$_4$]
molecular emission of 139-409 (left) and 134-411 (right)
computed along the directions with PA$=$87$^\circ$ and PA$=$115$^\circ$, 
respectively.
The velocity and angular resolutions are 4 km s$^{-1}$ and $\sim$ 1$''$, 
respectively.
The horizontal lines indicate the position of the center of each source,
while the vertical lines indicate the systemic velocity of the cloud.
The systemic LSR radial velocity of the ambient molecular cloud is about 7 
km s$^{-1}$.
Note the velocity gradients shown by the white lines.  In the top 
image the contours
are 2, 3, 4, 5, 6, 7, 8, 9, 10, 12, 14, 16, 18, 20, 22, and 24 times 120 
mJy beam$^{-1}$, the rms
noise of the image. In the bottom image the contours are 2, 3, 4, 5, 
6, 7, 8, 9, and 10 times
120 mJy beam$^{-1}$, the rms noise of the image. Note 
that our spectral resolution is comparable with
the linewidth of the CH$_3$CN transition.}
\end{center}
\label{fig3}
\end{figure}

\subsection{Two Circumbinary Molecular Rings?}

Since the CH$_3$CN molecule is a high-density tracer (Cesaroni et al. 
1999), 
the elongated structures traced by this molecule in 139-409
and 134-411 suggest two flattened circumbinary molecular rings
observed in nearly edge-on.
Moreover, the first moment of the CH$_3$CN[12-11] emission
shows a total velocity gradient of about a few kilometers per second that  
seems
to be aligned with the major axes of both structures, supporting also
this interpretation (see Figure 2). We note that
the two 7mm sources are aligned along
the
velocity gradient in 139-409, but perpendicular to it in 134-411. One
might speculate that the former ring is almost edge-on, whereas the
latter could be significantly inclined with respect to the line of sight.
In fact, if the binary system is coplanar with the ring and seen
edge-on, the two stars must appear along the direction of the velocity
gradient; instead, if the ring is inclined, the projections of the two
stars on the plane of the sky may appear along any direction. The fact
that the velocity gradient is much less in 134-411 than in 139-409 is
consistent with this interpretation.
Finally, in Figure 3 we show that
the kinematics of the molecular gas in both
cores, as computed along the major axes, seems to be
consistent with a ``rigid body law''.
This kinematic behavior also suggests
that the molecular gas in both flattened structures seems
to be only in a rotating {\it ring}.

Those molecular ring-shaped circumbinary structures have been also 
observed in
low-mass young stars, e.g. GG Tau and UY Aur (Guilloteau, Dutrey, 
\& Simon 1999;
Duvert et al. 1998). This kind of structures appear to be tidally stimulated 
during the formation
of binary or multiple systems of stars at its center.
When a multiple system is formed the very compact circumstellar
disks associated with the protostars clean its vicinity (or the innermost 
part of the large
circumbinary disk) forming thus a circumbinary ring around them (Mathieu 
et al. 2000).
This circumbinary ring might not transfer material to the compact 
circumstellar disks anymore,
however, see Monin et al. (2006) for a more detail discussion of this 
phenomenon.

It is interesting to note that much larger molecular rotating 
structures have also been observed
around multiple OB (proto)stars (Torrelles et al. 1983;  Beltr\'an et al. 
2005; Sollins et al 2005).
However, contrary to these circumbinary rings, these molecular structures 
(also known as ``toroids'')
seem to be transient and in non-gravitational-equilibrium (Beltr\'an et 
al. 2005). 

In conclusion, we interpret the flattened molecular structures 
associated
with the sources 139-409 and 134-411 as two circumbinary rings 
around binary systems traced by very compact 7 mm circumstellar 
disks
associated with young intermediate-mass (proto)stars.
However, we believe that more observations with higher angular
and spectral resolution are necessary to confirm our interpretation. 
Finally, this result supports
also the idea that intermediate-mass stars form through 
circumstellar disks and jets/outflows,
as the low mass stars do.

We thank the anonymous referee for many valuable suggestions. 
LFR acknowledges the support of CONACyT, M\'exico and DGAPA, UNAM.

Facilities: {\it SMA and VLA}

\end{document}